\documentclass[prbr,aps,twocolumn,bibnotes,superscriptaddress,epsf,citeautoscript]{revtex4-1}

\usepackage{graphicx}
\usepackage{amsmath}
\usepackage{amssymb}
\usepackage{color}
\usepackage{dcolumn}
\usepackage{bm}

\begin{document}

\title{Evidence for $s$-wave pairing with atomic scale disorder in the van der Waals superconductor NaSn$_2$As$_2$}

\author{K. Ishihara}
\affiliation{Department of Advanced Materials Science, University of Tokyo, Kashiwa, Chiba 277-8561, Japan}

\author{T. Takenaka}
\affiliation{Department of Advanced Materials Science, University of Tokyo, Kashiwa, Chiba 277-8561, Japan}

\author{Y. Miao}
\affiliation{Department of Advanced Materials Science, University of Tokyo, Kashiwa, Chiba 277-8561, Japan}

\author{O. Tanaka}
\affiliation{Department of Advanced Materials Science, University of Tokyo, Kashiwa, Chiba 277-8561, Japan}

\author{Y. Mizukami}
\affiliation{Department of Advanced Materials Science, University of Tokyo, Kashiwa, Chiba 277-8561, Japan}

\author{H. Usui}
\affiliation{Department of Physics, Osaka University, Toyonaka, Osaka 560-0043, Japan}

\author{K. Kuroki}
\affiliation{Department of Physics, Osaka University, Toyonaka, Osaka 560-0043, Japan}

\author{M. Konczykowski}
\affiliation{Laboratoire des Solides Irradi{\'e}s, {\'E}cole Polytechnique, CNRS, CEA, Universit{\'e} Paris-Saclay, F-91128 Palaiseau, France}

\author{Y. Goto}
\affiliation{Department of Physics, Tokyo Metropolitan University, Hachioji, Tokyo 192-0397, Japan}

\author{Y. Mizuguchi}
\affiliation{Department of Physics, Tokyo Metropolitan University, Hachioji, Tokyo 192-0397, Japan}

\author{T. Shibauchi}
\affiliation{Department of Advanced Materials Science, University of Tokyo, Kashiwa, Chiba 277-8561, Japan}

\date{\today}

\begin{abstract}

The recent discovery of superconductivity in NaSn$_2$As$_2$ with a van der Waals layered structure raises immediate questions on its pairing mechanism and underlying electronic structure. Here, we present measurements of the temperature-dependent magnetic penetration depth $\lambda(T)$ in single crystals of NaSn$_2$As$_2$ down to $\sim40$ mK. We find a very long penetration depth $\lambda (0) = 960$ nm, which is strongly enhanced from the estimate of first-principles calculations. This enhancement comes from a short mean free path $\ell \approx 1.7$ nm, indicating atomic scale disorder possibly associated with the valence-skipping states of Sn. The temperature dependence of superfluid density is fully consistent with the conventional fully gapped $s$-wave state in the dirty limit. These results suggest that NaSn$_2$As$_2$ is an ideal material to study quantum phase fluctuations in strongly disordered superconductors with its controllable dimensionality. 

\end{abstract}

\maketitle

Very recently, the discovery of superconductivity at a transition temperature $T_c \sim 1.3$ K has been reported in NaSn$_2$As$_2$ \cite{Goto}, which turns out to be the first superconductor in tin-pnictide-based layered materials \cite{Goto2}. The crystal structure of NaSn$_2$As$_2$ consists of a honeycomb lattice of SnAs conducting layers and Na spacer layers, and such a honeycomb layered structure has some resemblance to those of intercalated graphite \cite{graphite}, HfNCl \cite{HfNCl}, and transition-metal dichalcogenides (TMDs) \cite{WTe2a,WTe2b,MoTe2,MoS2}. A remarkable aspect of the SnAs-based layered materials is that each Sn$_2$As$_2$ bilayer is formed by van der Waals (vdW) forces, and thus ultrathin crystals of NaSn$_2$As$_2$ with the thickness of a few nanometers can be obtained by mechanical exfoliation \cite{Arguilla}. This makes NaSn$_2$As$_2$ potentially suitable for investigating two-dimensional (2D) superconductivity. 

It should be also pointed out that some related compounds with Sn$_2$As$_2$ bilayers are also known to have intriguing properties. These include SrSn$_2$As$_2$, which is considered to be a three-dimensional Dirac semimetal \cite{Gibson,Rong}, and EuSn$_2$As$_2$ with an antiferromagnetically ordered state below $T_N = 24$ K \cite{Arguilla2}, which may provide an intriguing platform for studying 2D magnetism. By creating a heterostructure with these SnAs-based vdW materials, novel interfacial states may emerge in the epitaxial combination of a superconductor, topological material and antiferromagnet \cite{nasn2as2comment}. For these interesting applications, a detailed understanding of the bulk nature of superconductivity in NaSn$_2$As$_2$ is indispensable.

Previous angle-resolved photoemission spectroscopy (ARPES) experiments of NaSn$_2$As$_2$ \cite{Arguilla} have suggested a possible multiband electronic structure, which cannot be reproduced by simple first-principles band-structure calculations. The specific heat measurements reveal the bulk nature of superconductivity, but information on the superconducting gap structure, which is the key to the pairing mechanism, has not been obtained because of the lack of low-temperature measurements \cite{Goto}. It also remains an open question whether or not the possible multiband effect on  superconducting gap is significant.

In this Rapid Communication, we measure the temperature dependence of the magnetic penetration depth $\Delta \lambda (T) = \lambda (T) - \lambda (0)$ at low temperatures which directly reflects the low-energy quasiparticle excitations. We observe the exponential temperature dependence of $\Delta \lambda (T)$, which indicates that no gap nodes exist on the Fermi surface. From the thermodynamic relation, we find that the magnitude of the penetration depth is as large as $\lambda (0) = 960 \pm 120$ nm, which is much longer than the London penetration depth $\lambda_L = 44$ nm obtained from the analysis using density functional theory (DFT) calculations. We also find from a comparison between the band-structure calculations and the measured residual resistivity that the electronic mean free path is as short as $\ell = 1.7$ nm, which is only 4.2 times as long as the nearest Sn-Sn distance. This supports a scenario where the valence-skipping states of Sn atoms lead to an intrinsic disorder of the charge density. In addition, the full temperature dependence of the superfluid density is well reproduced with that of the BCS dirty limit, consistent with a short mean free path. Moreover, to see the possible multiband effect of superconductivity, we introduce point defects by electron irradiation and find that the fully gapped nodeless state is robust against impurity scattering. These results demonstrate sign-preserving $s$-wave symmetry with a single gap, which implies that conventional superconducting pairing prevails over the atomic scale disorder inherent in NaSn$_2$As$_2$.

Single crystals of NaSn$_2$As$_2$ were prepared by a melt synthesis of raw elements, Na, Sn, and As with $1 : 2 : 2$ stoichiometric loading, and characterized by x-ray diffraction (XRD) and energy-dispersive x-ray spectroscopy \cite{Goto}. The samples were cut into typical dimensions of about $350 \times 350$ $\mu$m$^2$ in the $ab$ plane and thickness up to 50 $\mu$m. To measure $\Delta \lambda (T)$, we use a tunnel diode oscillator (TDO) operating at 13.6 MHz in a dilution refrigerator, and the measurements were performed down to $\sim 40$ mK. The resonant frequency shift of TDO $\Delta f$ relates linearly to the change of the magnetic penetration depth, $\Delta \lambda (T) = G \Delta f (T)$, where the geometric factor $G$ is determined by the geometry of the sample and the coil \cite{Carrington,Prozorov}. We calculated a demagnetization factor $N$ by using $N_{\rm {approx}} = 4L_a L_b / [4L_a L_b + 3L_c (L_a + L_b)]$ and $N-N_{\rm {approx}}$ calculated in Ref. \cite{demag}, where $L_a$, $L_b$, and $L_c$ are the rectangular sample dimensions and a weak ac magnetic field is applied parallel to the $c$ axis. Electron irradiation is used to study the effects of impurity scattering introduced by point defects  \cite{Mizukami,Cho}. We use the SIRIUS Pelletron linear accelerator operated by the Laboratoire des Solides Irradi{\'e}s (LSI) at {\'E}cole Polytechnique, and electrons with an incident energy of 2.5 MeV are irradiated at $\sim$20 K. 

\begin{figure}[t]
\includegraphics[keepaspectratio=true,clip,width=80truemm]{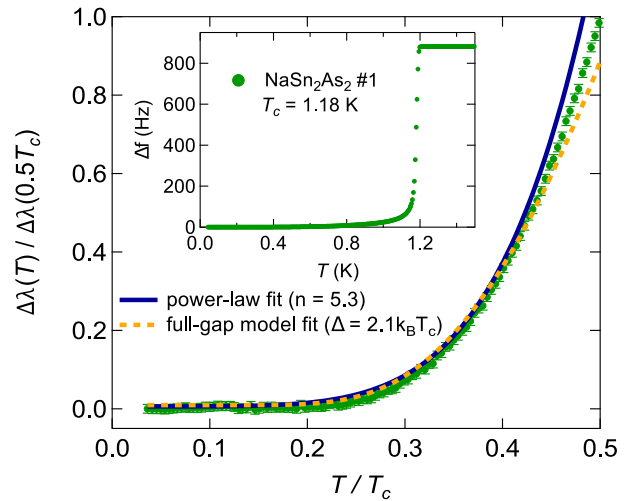}
\caption{Inset: The temperature dependence of resonant frequency shift in a range of $T < 1.5$ K for a single crystal of NaSn$_2$As$_2$ measured with the TDO system. The transition temperature $T_c = 1.18$ K is determined as the midpoint of the total shift. Main panel: The change of the magnetic penetration depth $\Delta \lambda (T)$ normalized by $\Delta \lambda (0.5 T_c )$ as a function of normalized temperature $T / T_c$. The blue solid and yellow dashed lines represent the fitting curves with power-law and fully gapped functions, respectively.}
\label{fig1}
\end{figure}

The inset of Fig. \ref{fig1} shows the temperature dependence of the resonant frequency shift for a sample whose surfaces were mechanically cleaved before the measurements. The observed sharp superconducting transition at $T_c \sim 1.18$ K indicates that the single crystal is chemically homogeneous. We note that no frequency shift is observed at the transition temperature ($\sim 3.7$ K) of Sn. The changes in the magnetic penetration depth $\Delta \lambda (T)$ at low temperatures is shown in the main panel of Fig. \ref{fig1}. To analyze these $\Delta \lambda (T)$ data, we used a power-law fitting, $\Delta \lambda \propto T^n$, in the range of $T < 0.4 T_c$, resulting in the exponent $n = 5.3$ (blue solid line in Fig.\,\ref{fig1}) that is much greater than the expected value of the nodal gap structure, $1 \leq n \leq 2$. In general, the power-law dependence with $n > 3$ is not distinguishable with exponential dependence. Therefore, we performed a full-gap model fitting, $\Delta \lambda (T) \propto T^{-1/2}\exp(-\Delta_{\mathrm {min}}/k_B T)$, where $\Delta_{\mathrm {min}}$ is the variable minimum gap. The calculated fitting parameter is $\Delta_{\mathrm {min}} = 2.1k_B T_c$ (yellow dashed line in Fig. \ref{fig1}), which is slightly larger than the value expected in the BCS theory ($\Delta = 1.76k_B T_c$), and this result is consistent with the value of the specific heat jump reported in previous studies \cite{Goto}. The observed exponential behavior of $\Delta \lambda (T)$ provides strong evidence for a nodeless gap structure of NaSn$_2$As$_2$.

\begin{figure}[t]
\includegraphics[keepaspectratio=true,clip,width=80truemm]{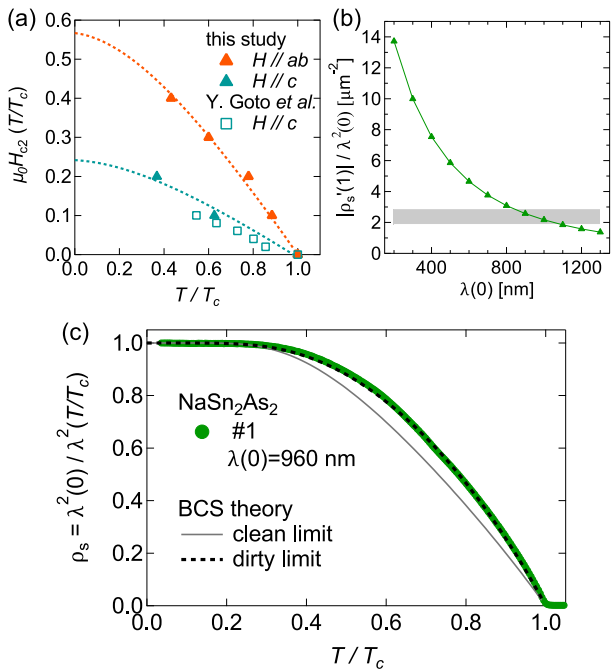}
\caption{(a) Upper critical field determined by zero resistivity with applied magnetic fields in the $ab$ plane (orange triangles) and along the $c$ axis (cyan triangles). The dashed lines represent approximate WHH curves for each data. For comparison, the zero resistivity $H_{c2}$ for $H \parallel c$ estimated from the data from Ref. \cite{Goto} is also plotted (cyan squares). (b) The green symbols are the values of $|\rho'_s (1)| / \lambda^2 (0)$ as a function of $\lambda (0)$. The gray band shows a range for the right-hand side of Eq. (1) with corresponding errors. (c) The green dots represent the superfluid density $\rho_s (T)$ calculated with $\lambda (0) = 960$ nm. The black dashed lines are the theoretical curves of single-gap BCS superconductors in the clean limit, and the gray solid line is that in the dirty limit.}
\label{fig2}
\end{figure}

The temperature dependence of normalized superfluid density, $\rho_s (T) = \lambda^2 (0) / \lambda^2 (T) = \lambda^2 (0) / (\lambda (0) + \Delta \lambda (T))^2$, provides more information on the superconducting gap function, such as gap anisotropy, scattering rate, and so on. To estimate $\lambda (0)$, here we use a relation between thermodynamic quantities known as the Rutgers formula \cite{Rutgers_formula},
\begin{equation}
\frac{\rho'_s (t=1)}{\lambda^2 (0)} = \frac{4\pi T_c \Delta C}{\phi_0 H'_{c2} (t=1)},
\label{Rutgers_eq}
\end{equation}
which is derived from Ginzburg-Landau (GL) theory and known to hold even in multiband superconductors \cite{Rutgers}.  Here $\phi_0$ is the magnetic flux quantum, $\Delta C$ is the specific heat jump at $T_c$, and $H_{c2}$ is the upper critical field, and the prime denotes a derivative with respect to the normalized temperature, $t = T / T_c$. By using the value of $\Delta C = 7.82$ mJ$\cdot$mol$^{-1}\cdot$K$^{-1}$ \cite{Goto} and $|\mu_0 dH_{c2} / dT|_{T_c} = 0.33$ T$\cdot$K$^{-1}$ [Fig. \ref{fig2}(a)], we obtain $|\rho'_s (1)| / \lambda^2 (0) = 2.38 \pm 0.51$ $\mu$m$^{-2}$ [gray band in Fig. \ref{fig2}(b)] with a $\pm 10$\% hypothetical error in  $|dH_{c2} / dT|$ and a $\pm 20$\% error in the $G$ factor. The plot of $|\rho'_s (1)| / \lambda^2 (0)$ as a function of $\lambda (0)$ is shown as the green symbols in Fig. \ref{fig2}(b), which provides $\lambda (0) = 960 \pm 120$ nm. We note that $\rho'_s (1)$ is determined in the range of $0.8 < T/T_c < 0.95$, which is considered to work well \cite{Rutgers}.

The full temperature dependence of $\rho_s (T)$ extracted by using the determined $\lambda (0) = 960$\,nm is shown in Fig. \ref{fig2}(c). The theoretical curves for single-gap BCS superconductors in the clean and dirty limits are also illustrated as gray solid and black dashed lines in the same figure, respectively. The curve of the BCS dirty limit reproduces very well the experimental data. Here, we point out that it is impossible to estimate the precise gap size of dirty superconductors from the $\rho_s(T)$ curve, because the curve in the dirty limit hardly changes with the gap size. If we use the clean limit, a gap size of $1.9 k_B T_c$ results in a similar temperature dependence as the BCS dirty limit. An important observation is that the $\rho_s (T)$ gives no evidence for multigap behavior, which would manifest itself as deviations from the single-gap temperature dependence or as a convex down curvature near $T_c$ \cite{Hashimoto}. 

\begin{figure}[t]
\includegraphics[keepaspectratio=true,clip,width=85truemm]{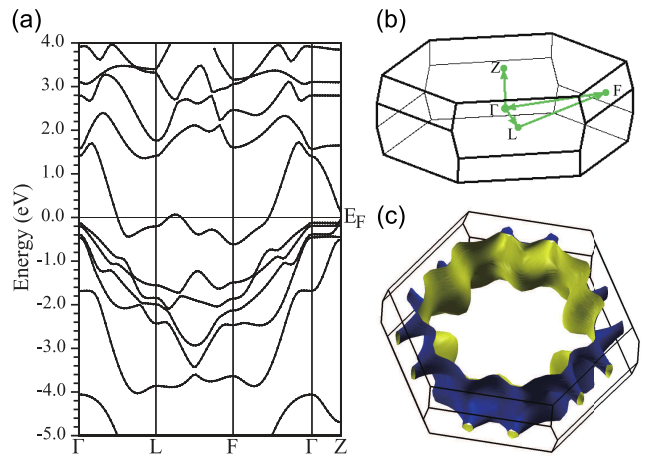}
\caption{(a) The electronic band structure of NaSn$_2$As$_2$ obtained by the DFT calculations. (b) Brillouin zone of rhombohedral lattice and high symmetry points. (c) The calculated Fermi surface of NaSn$_2$As$_2$.}
\label{fig3}
\end{figure}

The determined value of $\lambda (0) = 960$ nm is much longer than that of most $s$-wave superconductors, typically $\lambda (0) < 100$ nm \cite{Al,Rutgers}. To estimate the London penetration depth $\lambda_L$ expected from the electronic structure in the clean limit, we performed DFT calculations with spin-orbit coupling using the package WIEN2K \cite{Blaha} and generalized gradient approximation (GGA) in the form of Perdew-Burke-Ernzerhof. $RK_{\rm {max}}$ was set to 7, and we took 1000 $k$ points for the self-consistent calculation and $2 \times 10^5$ $k$ points for the calculation of the density of states, the carrier density and the Fermi velocity. The carrier density and the Fermi velocity are evaluated by using the BOLTZTRAP code \cite{Madsen2006}. Figure \ref{fig3}(a) shows the band dispersion in lines connecting between the high symmetry points of the rhombohedral lattice systems illustrated in Fig. \ref{fig3}(b) and the obtained Fermi surface is shown in Fig.\,\ref{fig3}(c). While the Fermi surface of ideal 2D materials has a cylindrical shape, the obtained Fermi surface of NaSn$_2$As$_2$ shows the existence of some warping in the Brillouin zone, which suggests the anisotropy of the bulk transport phenomena is not strong regardless of its vdW structure. This weak anisotropy is also confirmed by measurements of the upper critical fields $H_{c2}^{ab}$ and $H_{c2}^c$ [Fig. \ref{fig2}(a)] determined by the temperature at which the resistivity becomes zero with magnetic fields applied parallel to the $ab$ plane and $c$ axis, respectively. The dashed lines in Fig. \ref{fig2}(a) show fitting curves by using the function $H_{c2} \propto 1-t-0.153(1-t)^2 -0.152(1-t)^4$, which is considered as a good approximation of Werthamer-Helfand-Hohenberg (WHH) theory in the dirty limit \cite{WHH,WHHapprox}. The experimental anisotropy parameter $\gamma$ is obtained as $1.5 \leq \gamma = H_{c2}^{ab} / H_{c2}^{c} \leq 2.5$, and this value is consistent with the theoretically calculated value $\gamma = \sqrt{\langle v_{ab}^2 \rangle /\langle v_{c}^2 \rangle} \approx 1.5$, where $v_{ab}$ and $v_{c}$ are Fermi velocities in the $ab$ plane and along the $c$ axis, respectively. This consistency implies that the DFT calculations capture the salient features of the electronic structure in NaSn$_2$As$_2$.

In clean superconductors, $\lambda (0)$ is given by the equation $1/\lambda^2 (0) = \mu_0 n_s e^2 / m^\ast$, where $n_s$ is the superconducting electron density and $m^\ast$ is the effective mass of carriers. Therefore the long $\lambda (0)$ leads to the possibilities of a low carrier density \cite{srbi2se3,yptbi,srtio3} and/or a heavy effective mass \cite{uemura,Bauer,takenaka}. In contrast to these possibilities, the carrier density and the effective mass in the calculations are estimated as $n = 9.0 \times 10^{21}$ cm$^{-3}$ and $m^\ast = 0.68 m_e$, respectively. These obtained values indicate that the experimentally determined magnetic penetration depth $\lambda_{\rm {eff}}$ is effectively enhanced from the London penetration depth $\lambda_L$ following the equation $\lambda_{\rm {eff}}/\lambda_L \approx \sqrt{\xi_0 / \ell}$ in the case of dirty superconductors ($\ell \ll \xi_0$). Here, $\xi_0$ is the Pippard coherence length and $\ell$ is the mean free path. For the estimations of the theoretical Sommerfeld coefficient and $\lambda_L$, we calculated the density of states (DOS) and the plasma frequency $\omega_p$. The calculated Sommerfeld coefficient $\gamma = 4.0$ mJ$\cdot$mol$^{-1} \cdot$K$^{-2}$ is very close to the experimental value $\gamma = 3.97$ mJ$\cdot$mol$^{-1} \cdot$K$^{-2}$ \cite{Goto}, indicating no significant mass enhancement, namely, electron correlations are weak in this material. The $\lambda_L$ can be calculated from the theoretical plasma frequency $\omega_p = 4.5$ eV by using $\lambda_L = c/\omega_p$, and the determined value is $\lambda_L = 44$ nm. This result is consistent with the dirty-limit behavior of $\rho_s (T)$ and the ratio $\ell/\xi_0 \approx 0.002$ is obtained.

The value of $\ell$ can be estimated by the following two independent ways. First, we use the equation $\ell = v_F^{ab} \tau = \hbar k_F /\rho_0 n e^2$, where $\tau$, $\rho_0$, $v_F^{ab}$, and $k_F$ represent the relaxation time, residual resistivity, Fermi velocity in the $ab$ plane, and the Fermi wave number, respectively. Using the values of measured $\rho_0 = 122$ ${\mu \Omega} \cdot$cm and calculated $k_F \approx 0.5$\,$\rm{\AA^{-1}}$, we obtain $\ell \approx 1.7$ nm. Second, we use the equation for the GL coherent length in the dirty limit, $\xi_{\rm {GL}} = \sqrt{\xi_0 \ell}$, and $\ell/\xi_0 \approx 0.002$ obtained above. The value of $\xi_{\rm {GL}}$ can be calculated from the experimental $H_{c2}^c$ shown in Fig.\ref{fig2}(a) as $\xi_{\rm {GL}} = \sqrt{\phi_0 /2\pi \mu_0 H_{c2}^c} =38$ nm, which also leads to $\ell \approx 1.7$ nm. Here we emphasize that this very short mean free path consistently obtained in the two independent procedures is only 4.2 times as long as the nearest Sn-Sn distance (0.4 nm), which immediately indicates the presence of atomic scale disorder. 

As a clear powder XRD pattern \cite{Goto} and a sharp superconducting transition are observed, some extrinsic scattering of impurity phases is unlikely to be the origin of the atomic scale disorder. Instead, we propose to take the valence-skipping states of Sn atoms into account, which may be responsible for an intrinsic disorder of charge density. Among SnAs-based vdW material systems $A$Sn$_2$As$_2$ ($A=$ Na, Sr, Eu), NaSn$_2$As$_2$ can be treated as a hole-doped system to Zintl phases, because the electronegativities of Na and As atoms are largely different, leading to a complete charge transfer and ionic bondings. In this perspective, the nominal valence of Sn is expected as 2.5+, which consists of Sn$^{2+}$ and Sn$^{4+}$ ions with a ratio of 3 : 1. In valence-skipping superconductors, such as Ba$_{1-x}$K$_x$BiO$_3$, Pb$_{1-x}$Tl$_x$Te, and AgSnSe$_2$, electronic mean free paths of less than 3.5 nm have been estimated \cite{bkbo,pbte,agsnse2}, and strong scattering is considered to be a result of intrinsic disorder due to the valence-skipping states. Therefore the short mean free path and associated strong scattering in NaSn$_2$As$_2$ may be also accounted for by the coexistence of Sn$^{2+}$ and Sn$^{4+}$. On the other hand, it is claimed that NaSn$_2$As$_2$ has a smaller number of lone pairs around the Sn atoms than NaSnAs and NaSnP from thermal conductivity measurements and an analysis of the spatial distribution of the difference charge density \cite{ZLin}. Since Sn is a valence-skipping element, the smaller number of lone pairs suggests the coexistence of different valence states of the Sn atoms, which is consistent with our proposal. We note that, while M\"{o}ssbauer spectroscopy suggests only one valence state of the Sn atoms \cite{mossbauer}, the valence states of Sn may fluctuate faster than the time scale of the M\"{o}ssbauer measurements \cite{xps}. More direct determinations of the Sn valence states deserve further studies with microscopic probes.

\begin{figure}[t]
\includegraphics[keepaspectratio=true,clip,width=80truemm]{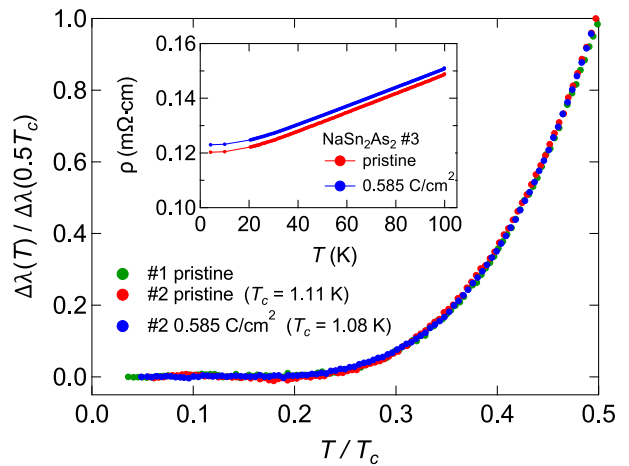}
\caption{(color online) Inset: Temperature dependence of resistivity, $\rho (T)$, of the pristine and irradiated samples (with a dose of  0.585\,C/cm$^{2}$). Main panel: The green, red, and blue symbols represent $\Delta \lambda (T) / \Delta \lambda (0.5 T_c )$ of pristine and irradiated samples with different $T_c$ values (1.18, 1.11, 1.08\,K), respectively.}
\label{fig4}
\end{figure}

In contrast to the DFT calculations, ARPES measurements report an additional small pocket near the $\Gamma$ point, indicating a multiband structure in NaSn$_2$As$_2$. This difference can be caused by charged surface states in ARPES measurements or unconsidered mixed valence states of Sn in DFT calculations, and the details of the correct band structure are still unsettled. We also note that the Hall coefficient of NaSn$_2$As$_2$ reported in Ref. \cite{Arguilla} depends significantly on temperature. The origin of this temperature dependence can be considered as a multiband effect or strong anisotropic scatterings, and thus it is not suitable to estimate the carrier density from the Hall effect measurements. However, if we take the Hall number at low temperatures as the carrier concentration, we find a factor of $\sim2$ shorter $\ell$ and thus our arguments discussed above are unchanged.

Provided that NaSn$_2$As$_2$ is a multiband superconductor, it is very important to determine the signs of the superconducting gap function in each band to elucidate the pairing mechanism. Therefore we performed electron irradiation and introduced point defects to see the change in the low-energy excitations. The resistivity curve $\rho (T)$ of the irradiated sample with a dose of 0.585 C/cm$^{2}$ shows a parallel shift behavior with increased $\rho_0$ values from that of the pristine sample, which demonstrates the enhancement of impurity scattering with little change in the band structure and inelastic scattering [the inset of Fig. \ref{fig4}(a)]. The penetration depth data in Fig. \ref{fig4}(a) show the robustness of full-gap $s$-wave superconductivity against impurity scattering in NaSn$_2$As$_2$.

The quantity of the induced point defects can be estimated with the pair-breaking parameter $g \equiv \hbar / \tau_{\mathrm {imp}} k_B T_{c0}$, where $\tau_{\mathrm {imp}}$ is the impurity scattering time and $T_{c0}$ is the transition temperature in a clean limit, and $\tau_{\mathrm {imp}}$ can be obtained using the relation $\tau_{\mathrm {imp}} = (\mu_0 \lambda_L^2 / \Delta \rho_0 )$, where $ \Delta \rho_0$ is the change of residual resistivity. By using the determined $\lambda_L = 44$ nm and $T_{c0} = 1.2$ K, the pair-breaking parameter increased by irradiation is calculated as $g = 78$. Comparisons with the results for a sign-changing $s_{\pm}$-wave superconductor BaFe$_2$(As$_{1-x}$P$_x$)$_2$ \cite{Mizukami}, and a $d$-wave superconductor, Ce$_{1-x}$La$_x$CoIn$_5$ \cite{115}, where the temperature dependence of $\Delta \lambda (T)$ changes into $\propto T^2$ due to impurity-induced low-energy Andreev bound states in the regime of $g > 3$, clearly indicate the robust exponential behavior of $\Delta \lambda (T)$ against point defects in  NaSn$_2$As$_2$, evidencing the sign-preserving $s$-wave symmetry of the gap function.

In summary, from magnetic penetration depth measurements on the pristine and irradiated samples of NaSn$_2$As$_2$, we find the robust exponential behavior of $\Delta \lambda (T)$ establishing nodeless $s$-wave symmetry of the superconducting gap function without a sign change. Both the value of $\lambda(0) = 960$ nm determined from Rutgers formula, which is much longer than the theoretical $\lambda_L = 44$ nm, and the $\rho_s (T)$ curve consistently indicate that the system is in the dirty limit ($\ell / \xi_0 \approx 0.002$). The analysis with comparisons with DFT calculations reveals the existence of atomic scale scattering with $\ell \approx 1.7$ nm which may be caused by intrinsic disorder of the charge density due to the mixed valence states of Sn atoms. Considering that strongly disordered superconductors are focused on in the context of quantum phase fluctuations and preformed Cooper pairs near the superconductor-insulator transition in 2D systems \cite{sittheory,honeycomb,sitexp} and a superconductor-metal transition in 3D systems \cite{agsnse2,nbn}, the vdW superconductor NaSn$_2$As$_2$, with the capability of mechanical exfoliation, provides a platform to study the quantum phase fluctuations and crossover of these anomalous transitions by the control of its dimensionality.

{\it Note added:} Recently, we became aware of a low-temperature thermal conductivity study of NaSn$_2$As$_2$, which consistently shows a fully gapped state \cite{Cheng}.

We thank the SIRIUS team, O. Cavani and B. Boizot, for running electron irradiation at LSI in {\'E}cole Polytechnique (supported by EMIR network, Proposal No. 16-9513). This work was supported by Grants-in-Aid for Scientific Research (KAKENHI) (Grants No. 15H02106, No. 15H05886, No. 15K17692, and No. 16H04493) from Japan Society for the Promotion of Science (JSPS).

\bibliography{refs}

\end{document}